# Postural instability in a young dyslexic adult improved by Hebbian pulse-width modulated lighting

Albert Le Floch [a,b,\*,] Samuel Henriat [c], Rosane Fourage [d], Guy Ropars [a,\*]

[a] Laboratoire de Physique des Lasers, UFR SPM, Université de Rennes 1, 35042 Rennes, France
[b] Laboratoire d'Electronique Quantique et Chiralités, 20 Square Marcel Bouget, 35700 Rennes, France
[c] Clinic of Podiatry, 56300 Pontivy, France
[d] Institut de Formation IFPEK, 35000 Rennes, France

**ABSTRACT**

Postural stability is linked to vision in everyone, since when the eyes are closed stability decreases by a factor of 2 or more. However, in persons with dyslexia postural stability is often deficient even when the eyes are open, since they show deficits in motor as well as specific cognitive functions. In dyslexics we have shown that abnormal symmetry between retinal Maxwell's centroid outlines occurs, perturbing the interhemispheric connections. We have also shown that pulse-width modulated lighting can compensate for this lack of asymmetry, improving the reading skills. As the postural stability and the vision are correlated, one may wonder if the excess of the postural instability recorded in a young adult with dyslexia can also be reduced by a pulse-width modulated light controlling the Hebbian synaptic plasticity. Using a force platform we compared the postural responses of an observer without dyslexia with the responses of a subject with dyslexia, by measuring their respective standing postures with eyes open looking at a target in a room with either continuous or pulse lighting. There was no effect of changing the lighting conditions on the postural control of the subject without dyslexia. However, we found that the postural stability of the subject with dyslexia which was actually impaired during continuous light, but was greatly improved when a 80 Hz pulsed light frequency was used. Importantly, the excursions of the surface area of the center of pressure on the force platform were reduced by a factor of 2.3.

\* Corresponding authors.
E-mail address: guy.ropars@univ-rennes1.fr (G. Ropars)
albert.lefloch@laposte.net (A. Le Floch)

# 1. Introduction

The role of vision in the maintenance of posture and the visual stabilization of posture has been observed for normal persons by many groups [1-4]. Postural sway measured using force platforms is commonly used to compare the human posture in quiet stance with eyes open and eyes closed for children and older people [5,6]. In these studies the postural fluctuations have been shown to increase when the eyes are closed, showing the important role of vision and neural connections in postural stability.

Developmental dyslexia is a complex specific learning difficulty that affects reading, writing, spelling and the development of literacy and language [7-9]. Moreover the problems faced by many dyslexic children are by no means confined to these difficulties [10]. The postural deficiency syndrome [11] and different impairments in motor skills [10,12], balancing ability [13], and postural control [14-18] are often observed in dyslexic people and in children with developmental disorders [19-20].

In dyslexic children and young adults, a whole-brain analysis of the connectivity using functional magnetic resonance has shown different disruptions of functional networks [21], with a lack of lateralization, namely in the left visual-word-form area [22] but also with highly significant differences in functional connectivity when compared to a control group. The use of the functional trans-cranial Doppler ultrasound method to directly access the cerebral asymmetry, has confirmed the reduction of the usual left-sided language lateralization in dyslexia [23]. The brain lateralization and the interhemispheric connectivity being closely linked to motor control and adaptation [24,25], impairments in the connectivity in dyslexia result in different possible motor impairments including motor overflow [26-29], posture instability [14-18], balancing ability in children [13,30] and in adults [31,32,33]. Indeed, brain activation during maintenance of standing postures in humans confirms that the visual cortex plays an important role [6,34]. The links between the primary visual and motor areas are complex and involve a series of cortico-cortical links. Moreover visual areas through the brain are connected to motor areas through the cerebellum which is strongly engaged in motor control including posture and balancing ability, while also playing a role in different cognition functions [35-38]. Unintentional motor impairments such as motor overflow with mirror movements of the hands and effects replicating primary reflex movements on specific reading difficulties in dyslexic children [39], also suggest weak cerebral asymmetry and weak brain lateralization in dyslexic people.

The lack of interhemispheric asymmetry linked to the Maxwell's centroid asymmetry in many dyslexic observers, has been shown to often include the simultaneous observation of veridical and mirror-images of letters like "b" and "d" [40]. Since this study with students, we have observed similar results with many children and teenagers from 8 to 17 years old. Moreover, as a pulse-width modulation LED lamp, using Hebbian mechanisms, is able to erase the mirror-images, restoring the reading skills and providing the person with dyslexia with an assisted cerebral asymmetry [40], one may wonder if the muscular postural instabilities which often accompany the reading difficulties in these dyslexic observers, can also be reduced. It is the aim of this paper to investigate the effect of a pulse-width light modulation on the postural stabilization of a dyslexic observer. We compare the standing postural responses in binocular fixation tests of a dyslexic observer to those of a control observer without dyslexia, under first continuous light regime and then pulsed light regime using Hebbian mechanisms.

# 2. Material and methods

## 2.1. Participants

The two volunteer male students (22 years old) take the same scientific course in our University. They were aware of the purpose of the study. The informed consent was obtained from each participant after explaining the nature of the study. One of the students, without dyslexia, has no specific reading problem, while the second student with dyslexia is assisted by the Health Center of the University which provides extra-time for the examinations. Both students have normal binocular vision. The entire investigation process was conducted according to the principles expressed in the Declaration of Helsinki.

## 2.2. Optical tests

### 2.2.1. *The Asymmetry of Maxwell's centroids*

Posture being linked to vision, we have first to investigate the presence or absence of imbalance in human vision, previously suspected by Hubel and Wiesel [41,42]. An asymmetry has been detected at the center of the two foveas by comparing the two small blue cone-free areas which correspond to the so-called Maxwell centroids [40]. As the retinas are parts of the central nervous system, this asymmetry plays a crucial role in the connectivity and the lateralization of the brain. For each eye, the blue cone-free area is seen as a small dark zone on a blue background when fixating a bright white screen through a blue filter. Using a blue-green exchange filter in a so-called foveascope [40], the contrast can be optimized and the outlines of the Maxwell's centroid entoptic images recorded for each eye. To quantify the ellipticity of each blue cone-free area, we use an osculating ellipse and define the ellipticity $\varepsilon_R$ and $\varepsilon_L$ for the right and the left eye respectively. The ellipticity difference $\Delta\varepsilon = \varepsilon_R - \varepsilon_L$ defines the Maxwell's centroid asymmetry for each observer, and the sign of this difference determines the eye dominancy which is confirmed by the afterimage test (see below).

2.2.2. *Noise-activated afterimages*

Eye dominancy being a hallmark of asymmetry in human vision, we had to determine the eye preference of the two observers. To avoid possible artefacts with the different sighting methods such as the distance of observations or the gaze angle, we use the noise-activated afterimage method [40], where the two eyes remain closed after a fixation by the two eyes on the same bright stimulus. The observer can look, for instance, at a black letter "c" on a bright screen. With his two eyes, the observer gazes at the "c" letter for about 10 seconds, then he closes his eyes and with his hands, blocks the 2 to 3 % [40] of the diffused light passing through his closed eyelids which can act as noise. The two retinas are then equally imprinted. Now, by alternatively shifting the hands in front of each eye with a periodicity of about 2 s, two noise-activated afterimages are seen successively, one for each eye. Hence, we have a differential method for comparing the activation in the brain of the same information encoded through each retina. The dominant eye corresponds directly to the brightest afterimage, i.e. to the eye with the strongest pathway to the brain.

The noise-activated afterimages have another interest. Indeed for many persons with dyslexia, the method allows us to directly see the mirror-image of letters which induces the crowding effect, detrimental to their reading skills. With the noise-activated afterimage test, the presence or the absence of extra mirror-images is easily detected. Moreover the technique allows us to directly observe the erasing of the mirror images using the Hebbian mechanisms in the case of a person with dyslexia, when they use a pulse-width modulation system during the fixation.

2.3. Posture measurements

The performances of the two observers were recorded by a force platform (Médicapteurs) following the standard instructions [43-46]. To record the movement of the body's center of pressure with both eyes open, there is a circular target, i.e. a red laser spot of about 5 mm diameter on a wall that was situated 90 cm in front of the subject at eye level. The room had a normal level of illumination. However the lighting was provided by a LED panel which was electronically controlled. The system can work either in the usual continuous lighting regime (Fig. 1A) or in a pulse-width modulation regime (Fig. 1B), with a repetitive frequency which can be varied from 65 to 120 Hz [40].

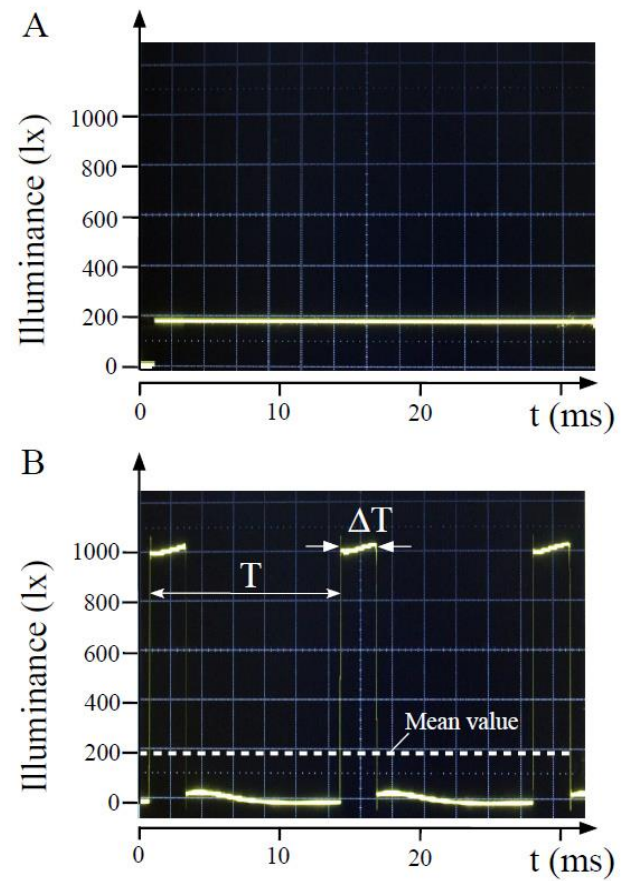

*Fig. 1. The two lighting regimes. A) Target illuminance in the continuous regime. B) Target illuminance in the 80 Hz pulsed regime, with a duty cycle $\Delta T/T = 0.2$, to use the Hebbian mechanisms. The average illuminance is 200 lx in the two cases.*

In the two regimes, the average illumination on the fixation target surface was 200 lx. The observers stood barefoot on the force platform in a position with feet abducted at 30°, heels separated by 5 cm, and their arms hanging loosely by their sides [17, 43-47]. The center of pressure sway signals and the latero-lateral and anterior-posterior signals were recorded with a sampling rate of 100 Hz for 30s trial recordings. Three trials are repeated for each situation. For each participant the binocular fixation was performed first under the continuous lighting regime, then under the pulsed regime.

## 3. Results

### 3.1. Optical tests

Figure 2 shows the standard vision recordings of the young adult without dyslexia.

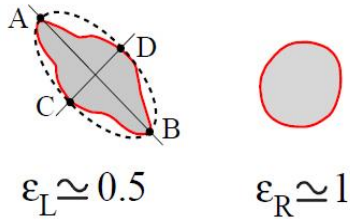

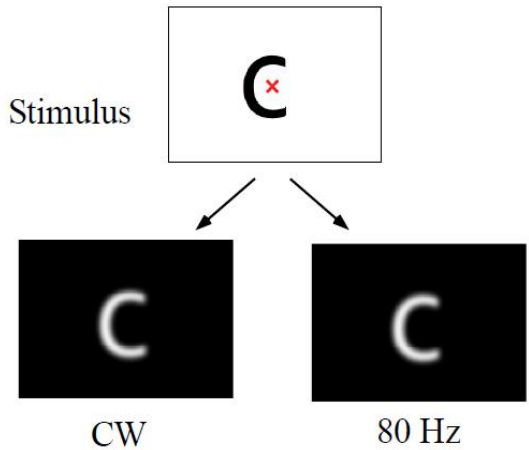

*Fig. 2. Maxwell's centroid outlines and afterimages for a young adult without dyslexia. A) The asymmetry of the Maxwell's centroid outlines: here $\Delta\varepsilon = \varepsilon_R - \varepsilon_L \approx + 0.5$. The quasi-circular right-eye outline for this observer corresponds to the dominant eye which is confirmed by the brighter noise-activated afterimage of his right eye. B) The observed afterimages of a "c" stimulus. The afterimages are unchanged with a continuous laboratory lighting or with an 80 Hz pulse-width modulated laboratory lighting.*

First, the outlines of the Maxwell's centroids he observed (Fig. 2A) show an asymmetry $\Delta\varepsilon = \varepsilon_R - \varepsilon_L \approx + 0.5$. The eye preference of this observer is fixed by the quasi-circular outline in his right eye. This eye preference is in agreement with the brighter noise-activated afterimage he sees with his right eye. Second, either in the continuous lighting regime or in the 80 Hz pulse lighting regime, the binocular afterimage of the "c" stimulus remains unchanged (Fig. 2B).

The corresponding results for the young adult with dyslexia are reported in Fig. 3.

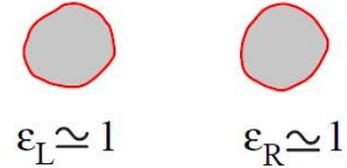

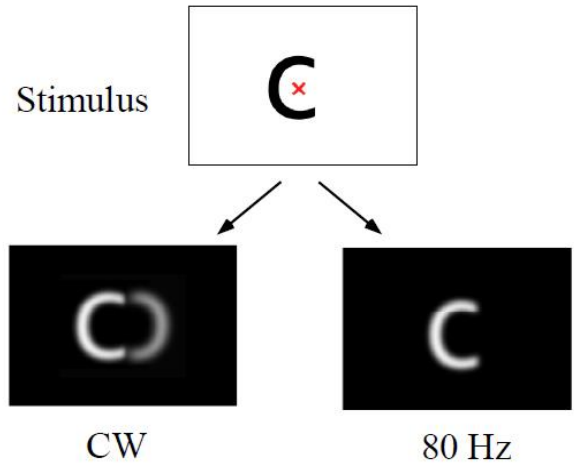

*Fig. 3. Maxwell's centroid outlines and afterimages for a young adult with dyslexia. A) The seeming symmetrical outlines. The two quasi-circular outlines correspond for this observer to their two similar noise-activated afterimages and induce a no-dominant eye status. B) The observed afterimages of a "c" stimulus. The afterimage in the continuous regime shows the coexistence of the primary image "c" and of the weaker mirror image "ɔ". In the 80 Hz pulsed regime, with a duty cycle of 0.2, the mirror-image is cancelled using the Hebbian mechanisms.*

Here, the asymmetry between the two Maxwell's centroids is $\Delta\varepsilon \approx 0$ (Fig. 3A). The two quasi-circular outlines correspond to an undetermined dominancy which is confirmed by the equal brightness of the afterimages seen by the two eyes of the observer. In contrast of the case of the non-dyslexic adult, the afterimage often seen by observers with dyslexia in the usual continuous lighting regime is the signature of a too high interhemispheric symmetry, i.e. a too weak lateralization in the brain [40]. Indeed here, the symmetric mirror-image of the stimulus "c" projected through the callosum between the two hemispheres is not weakened enough, and coexists with the veridical afterimage, perturbing the observer's reading skills (left part of Fig. 3B). However, as described in ref. 40, when the afterimages are observed using a 80 Hz pulse light during the fixation, the extra mirror-image is erased (right part of Fig. 3B), and the observed afterimage is identical to that seen by the observer without dyslexia. The Hebbian mechanisms governing the synapse plasticity contribute to the weakening of the mirror-image, induce an assisted asymmetry and a strengthened lateralization in the brain. The cognitive functions such as reading are improved and we can also now investigate the respective postural effects for the two observers in the continuous or pulsed light regimes.

3.2. Postural measurements

On Fig. 4A the latero-lateral and the anterior-posterior performances of the young adult without dyslexia were recorded for 30s trials.

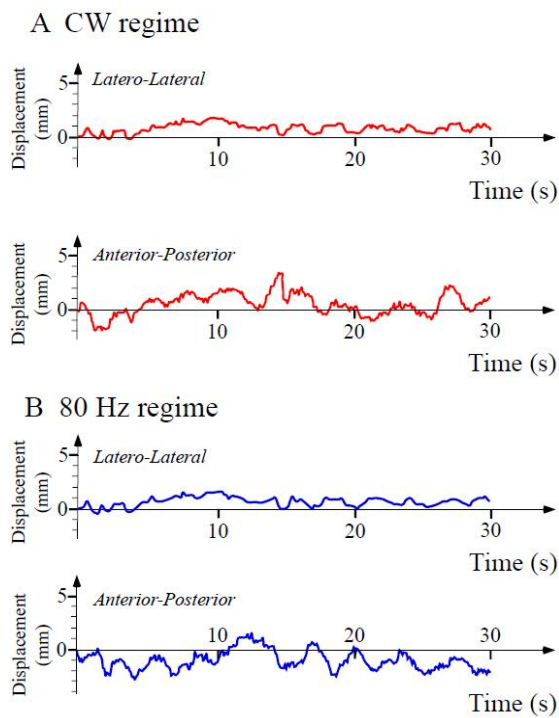

*Fig. 4. Force platform results for a young adult without dyslexia. A) Latero-lateral and anterior-posterior variations under CW laboratory lighting. The surface of the stabilogram is 10 $mm^2$. B) Same results under 80 Hz pulse laboratory lighting. The results are unchanged (here the surface of the stabilogram is 8 $mm^2$).*

Following the standard instructions three trials are repeated for each situation. The respective variations were rather weak. No appreciable changes were observed in the 80 Hz pulsed light regime. The surface of the center of pressure was $10 \pm 2$ $mm^2$ continuous regime, and $8 \pm 2$ $mm^2$ in the pulsed regime.

The corresponding measurements for the young adult with dyslexia are reported in Fig. 5.

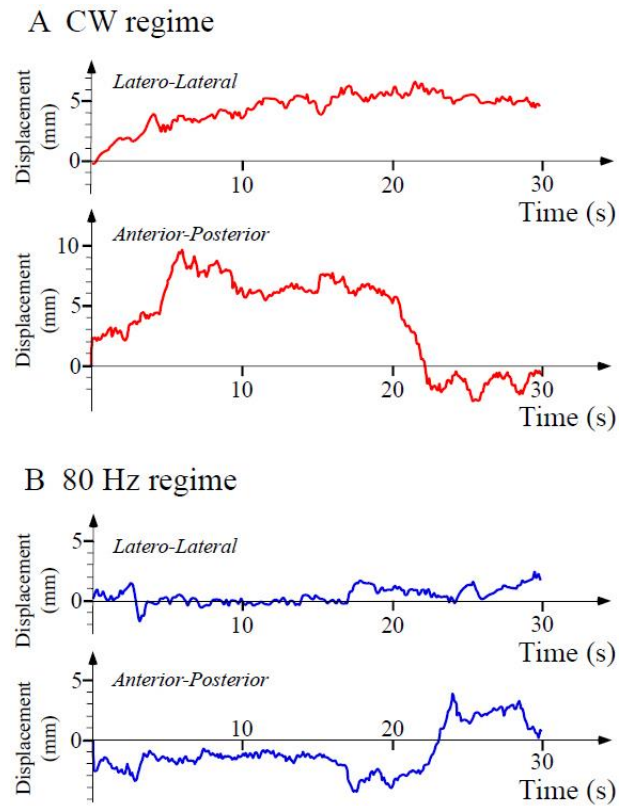

*Fig. 5. Force platform results for a young adult with dyslexia. A) Latero-lateral and anterior-posterior variations under continuous lighting. The results show a postural deficiency, which are in agreement with the 66 $mm^2$ surface for the centre of pressure. B) Same results under 80 Hz pulsed lighting. Both the latero-lateral and the anterior- posterior variations are reduced by a factor 1.5 and the surface (28 $mm^2$) is reduced by a factor 2.3 thanks to the Hebbian mechanisms, showing the neural plasticity.*

In the continuous lighting regime (Fig. 5A) the latero-lateral and anterior-posterior oscillations were significantly larger compared to those of the first observer, by approximately a factor 3. The surface area of the center of pressure reached $66 \pm 7$ $mm^2$. However, interestingly, in the presence of the 80 Hz pulsed lighting, the latero-lateral and anterior-posterior oscillations were reduced by a factor 1.5 and the surface area of the center of pressure ($28 \pm 4$ $mm^2$) was reduced by a factor 2.3. The same tests repeated six months later confirm the results. Moreover, according to the Hebbian mechanisms governing the synaptic plasticity [48], for a non-matched modulation frequency of the lighting like 120 Hz, the stability was not improved as at 80 Hz, but was worsened. The displacement variations were increased by a mean value of 1.25, and the surface was increased by a factor of 1.5 (Fig. 6). The stability measurements for the control observer remained unchanged irrespective of the lighting frequency.

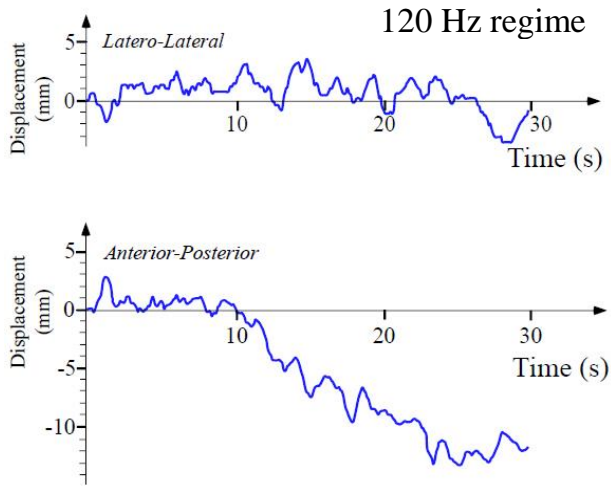

*Fig 6. Force platform results for the young adult with dyslexia under 120 Hz pulse lighting. Both the latero-lateral and the anterior-posterior variations are increased by a factor 1.3 and 1.2 respectively and the surface (100 $mm^2$) is increased by a factor 1.5, when compared to the continuous lighting corresponding here to the strengthening of the Hebbian synaptic plasticity. Note that the surface for the 120 Hz pulsed lighting is 3.5 times larger than the surface for the 80 Hz pulsed lighting.*

## 4. Discussion

The postural performances for a person with a normal condition, although strongly linked to vision, remained unchanged in binocular vision fixation of a target whatever the continuous or pulsed regime of the lighting. By contrast the postural instabilities which persist for the young adult with dyslexia, even with the eyes open, were shown to be controlled by the lighting regime which also controls specific cognitive functions.

The strengthening of the asymmetry and of the lateralization via the Hebbian mechanisms [40,48] exploiting the neural plasticity which permits the interhemispheric mirror-images in the reading tasks for a person with dyslexia to be erased, could also reduce the postural instabilities. This suggests that the visual information reaches the cerebellum which seems to be the most important link between visual and motor areas [49]. The brain cortex is known to be bidirectionally connected to the cerebellum through multiple circuits. Standing being a process of feedback control, involves continuous compensatory adjustments of the musculature. Indeed, neurons of the visual cortex communicate through extensive axonal projections targeting highly diverse areas [50], and posture cells have reciprocal connections to various cortical areas including the visual cortex [51-52]. These neuronal bonds are likely to be involved in our postural observations.

## 5. Conclusion

To conclude, vision and posture deficits appear intrinsically linked in the brain of the observer with dyslexia. The two retinas, as parts of the central nervous system, send a large mass of information to the cortex which processes the information to be transferred to other brain areas and the rest of the body. The stabilometric experiments with the young adult with dyslexia, suggest that the vision controls the posture. The lack of asymmetry in the young dyslexic adult can then be optically assisted via the Hebbian mechanisms governing the synaptic plasticity, improving both the specific cognitive functions like reading, and by cascading effects improving the motor functions like posture for a 80 Hz pulse modulated lighting. We hope that these postural results will be observed by other groups. As among the different sensory inputs to the brain vision seem to play a crucial role, the pulse-width modulated light effect could perhaps also be observed in the developmental coordination disorders such as dyspraxia and attention deficit hyperactivity disorder. Further studies will be needed for an extension to the different cases and perhaps will lead to useful applications.

**Ethics**
This study was conducted according to the principles expressed in the Declaration of Helsinki.

**Conflict of interest:**
None of the authors have a conflict of interest.

**Acknowledgments:**
The authors wish to thank the two students for their kind participation, and B. Mitchell and K. Dunseath for the proofreading of the manuscript.

**Funding**
We received no funding for this study.

**Data accessibility**
This article has no additional data.